\documentstyle [12pt]{article}
\makeatletter \oddsidemargin  0in \evensidemargin 0in
\textwidth16cm \RequirePackage[dvips]{graphicx} \textheight 20.5cm
\newcommand{\singlespacing}{\let\CS=\@currsize\renewcommand{\baselinestretch}{1.5}\tiny\CS}
\makeatletter \oddsidemargin  -.1in \evensidemargin -.1in
\newcommand{\doublespacing}{\let\CS=\@currsize\renewcommand{\baselinestretch}{1.35}\tiny\CS}

\def\@citex[#1]#2{\if@filesw\immediate\write\@auxout{\string\citation{#2}}\fi
  \def\@citea{}\@cite{\@for\@citeb:=#2\do
    {\@citea\def\@citea{,\linebreak[0]\hskip0pt plus .2em}%
      \@ifundefined{b@\@citeb}%
    {{\bf ?}\@warning{Citation `\@citeb' on page \thepage\space undefined}}%
      \hbox{\csname b@\@citeb\endcsname}}}{#1}}

\newtheorem{rule-def}[theorem]{Rule}
\begin{document}
\title{\bf Teleportation via a mixture of a two qubit subsystem of a N-qubit W and GHZ state}\author{I.Chakrabarty $^{1,2}$\thanks{Corresponding author:
E-Mail-indranilc@indiainfo.com }\\
$^1$ Heritage Institute of Technology, Kolkata-700107, West Bengal, India\\
$^2$ Institute of Physics, Sachivalaya Marg, Bhubaneswar-751005, Orissa, India }
\date{}
\maketitle{}
\begin{abstract}
In this work we study a state which is a random
mixture of a two  qubit subsystem of a $N$-qubit W state and GHZ
state. We analyze several possibilities like separability
criterion (Peres-Horodecki criterion [14,15]), non violation of
Bell's inequality [6]($M(\rho)<1$)and teleportation fidelity
[1,2,3,4] ($(F_{max}>\frac{2}{3})$) for this state. We also obtain a relationship between $N$ (number of qubits)
and $p$ (the classical probability of random mixture) for each of
these possibilities. Finally we present a detailed analysis of all these possibilities for $N=3,4,5$ qubit systems.  We also report that for $N=3$  and $p\in(.75,1]$, this entangled state can be used as a teleportation channel without
violating Bell's inequality.
\end{abstract}
\section{Introduction}
In the early nineties a group of six scientists discovered a new
aspect of quantum entanglement - teleportation. Teleportation is
purely based on classical information and non-classical
Einstein-Podolsky-Rosen (EPR) correlations. The basic scheme of
teleportation is to transfer an arbitrary quantum state from
sender to receiver using a pair of particles in a singlet state
shared by these two distant persons. It was interesting to find
out the optimal value of the teleportation fidelity of an unknown
quantum state; a fidelity above which will ensure us about the 
non-classical character of the state forming the quantum channel.
It has been shown that for a purely classical channel the optimum
teleportation fidelity is $F=\frac{2}{3}$ [2,3,4].\\
In reference [5] the author raised an important question
regarding quantum teleportation, Bell-CHSH inequalities [6] and
inseparability: ``What is the exact relation between Bell's
inequalities violation and teleportation?''. \\
Bell's inequalities are basically built upon the locality and
reality assumption and have nothing to do with quantum mechanics.
However, it is the fact that quantum mechanics predicts a
violation of these conditions that makes them interesting.
However in another work Werner  gave an example of an entangled
state, which has the unique feature of not violating Bell's
inequality for a certain range of the classical probability of mixing
[7]. At this point one can ask whether
states which violate Bell-CHSH inequalities are suitable for
teleportation or not. In another work [8], the authors addressed this
question in the form of a result that tells us that any mixed two
spin$\frac{1}{2}$ state violating the Bell-CHSH inequalities is
suitable for teleportation. However since the exact relationship between Bell’s
inequality and teleportation is unknown, it remains interesting to see whether there exists
any entangled state which does not violate Bell’s inequality but still can act as a
teleportation channel with a fidelity $>\frac{2}{3}$. In our recent work we have showed that the
output entangled state of Buzek-Hillery cloning machine satisfy Bell’s inequality and
at the same time can be used as a teleportation channel for certain range of the machine
parameter [9]. \\
True tripartite entanglement of the state of a system of three
qubits can be classified on the basis of stochastic local
operations and classical communications (SLOCC). Such states can
be classified in to two categories: GHZ states and W-states. It is
known that GHZ states can be used for teleportation and super
dense coding, but the prototype W-states cannot be. However in [10]
the same authors proved that the prototype W state of the form
$|W\rangle=\frac{1}{\sqrt{3}}[|001\rangle+|010\rangle+|100\rangle]$
cannot be used as a teleportation channel.  In another piece of
work it has been shown that if one uses W-states then the
teleportation protocol works with non-unit fidelity, i.e., it is
not perfect [11]. Also, it is not possible to recover the unknown
state using W-state as a channel . However in another work it has
been shown that there is a class of W-states that can be used
for perfect teleportation and super dense coding [12].\\
In this work, in the first section we study whether the two qubit
subsystem of a $N$ qubit W state of the form $|W\rangle=
\frac{1}{\sqrt{N}}{|100...0\rangle+|010...0\rangle+.....+|000...1\rangle}$
can be used as a teleportation channel or not. Our result shows that
the maximum fidelity of teleportation for such a state is
$\frac{2}{3}$. We also investigate the criterion like violation of Bell's inequality for this state [5].
In the next section we
study a new state by taking a classical mixture of a two qubit
subsystem of a $N$-qubit W state and a $N$-qubit GHZ state. This state is of
the form; $\rho_{GHZ,W}= pTr_{N-2}(|W\rangle\langle
W|)+(1-p)Tr_{N-2}(|GHZ \rangle\langle GHZ|)$; $(0<p<1)$. We investigate its entanglement properties and
efficiency as a teleportation channel. We further show
that this new state satisfies the Bell-CHSH inequality. In the next
section we present a detailed analysis for $N=3,4,5$. We find out that for $N=3$  and $p\in(.75,1]$, the state not only satisfies Bell's inequality and on the same time can also be used as a resource for teleportation.
 This work also generalizes a recent work
[13] where authors also considered a same kind of state for $N=3$.\\
In a nutshell, in this work we look out for an entangled state which
can act as a teleportation channel without violating Bell's
inequality. In that process we consider the state
$\rho_{GHZ,W}$ and study its characteristics. This state can also be viewed as a mixture of
separable and entangled states. The basic motivation for studying
such a specific state, is to analyze how such a classical mixture
of entangled and separable states plays a pivotal role in
teleportation process.
In other words, in this work we investigate how the classical
probability  of mixing $p$ affects phenomenon like violation of Bell's
inequality and quantities like teleportation fidelity.\\

\section{N-qubit W state : Bell's Inequality and Teleportation}
In this section we study the efficacy of a
two qubit subsystem of a $N$-qubit W state as a resource for quantum teleportation.
Further we investigate whether this state satisfies Bell-CHSH inequality or not.\\
We consider a $N$ qubit W state of the form
\begin{eqnarray}
|W\rangle=
\frac{1}{\sqrt{N}}{|100...0\rangle+|010...0\rangle+.....+|000...1\rangle}.
\end{eqnarray}
After tracing out any $(N-2)$ qubits, the reduced density matrix
representing its subsystem is given by,
\begin{eqnarray}
\rho_{ij}(i,j=1,...N)&=&tr_{N-2}(|W\rangle\langle
W|)=\frac{1}{N}[|01\rangle\langle 01 |+|10\rangle\langle 10
|{}\nonumber\\&&+|01\rangle\langle 10|+|10\rangle\langle
01|+(N-2)|00\rangle\langle 00|].
\end{eqnarray}
The necessary and sufficient condition for any state $\rho$ of
two spins $\frac{1}{2}$ to be inseparable is that at least one of
the eigen values of the partially transposed operator defined as
$\rho^{T_2}_{m\mu,n\nu}=\rho_{m\nu,n\mu}$, is negative [15]. This
is equivalent to the condition that at least one of the two
determinants\\
$W_{3}= \begin{array}{|ccc|}
  \rho_{00,00} & \rho_{01,00} & \rho_{00,10} \\
  \rho_{00,01} & \rho_{01,01} & \rho_{00,11} \\
  \rho_{10,00} & \rho_{11,00} & \rho_{10,10}
\end{array}$ and $W_{4}=\begin{array}{|cccc|}
   \rho_{00,00} & \rho_{01,00} & \rho_{00,10} & \rho_{01,10}\\
  \rho_{00,01} & \rho_{01,01} & \rho_{00,11} & \rho_{01,11} \\
  \rho_{10,00} & \rho_{11,00} & \rho_{10,10} & \rho_{11,10} \\
  \rho_{10,01} & \rho_{11,01} & \rho_{10,11} & \rho_{11,11}
\end{array}$\\
is negative [14,15].\\
Here we investigate the inseparability of the two qubit density
operator $\rho_{i,j}$ for different values of $N$. After
evaluating the determinants for $\rho_{ij}$, we obtain the values
of $W_3$ and $W_4$ as
\begin{eqnarray}
W_3=\frac{N-2}{N^3},~~~~W_4=-\frac{1}{N^4}
\end{eqnarray}
It is evident that $W_3>0~~~ (\forall N>2$) and $W_4<0~~~
(\forall N)$. Since we find that at least one of these two determinants is
less than zero for all values of $N$, so we conclude with full generality
the state $\rho_{ij}$ to be inseparable.\\
Now, it remains interesting to see  whether the reduced density matrix violates
Bell's inequality or not. It is a known fact that the state which
does not violate Bell's inequality must satisfy $M(\rho)\leq1$. The quantity $M(\rho)\leq1$ is defined by $M(\rho)=\max_{i>j}(u_i+u_j)$ [3], where $u_i$ and $u_j$ are
the eigenvalues of the matrix $U=C^t(\rho)C(\rho)$. The elements of the matrix  $C(\rho)=[C_{ij}]$ are given by 
$C_{ij}=Tr[\rho\sigma_i\otimes \sigma_j]$, where $\sigma_i$ are Pauli spin matrices [8].\\
The eigen values of the matrix $U=C^t(\rho_{ij})C(\rho_{ij})$ for
the bipartite state $\rho_{ij}$ are given by,\\
$u_1=u_2=\frac{4}{N^2}$, $u_3=\frac{(N-4)^2}{N^2}$.\\
Now we consider the following three cases separately:\\
\textbf{Case 1:} $u_3>u_1=u_2$, when $N>6$\\
Hence $M(\rho_{ij})= \frac{(N-4)^2}{N^2}+\frac{4}{N^2}<1$.\\
\textbf{Case 2:} $u_1=u_2>u_3$ when $3\leq N<6$\\
Hence $M(\rho_{ij})=
\frac{4}{N^2}+\frac{4}{N^2}=\frac{8}{N^2}<1$.\\
\textbf{Case 3:} $u_1=u_2=u_3$ when $N=6$ \\
Hence $M(\rho_{ij})=\frac{4}{N^2}+\frac{4}{N^2}=\frac{8}{N^2}<1$.\\
We see that in each of the three cases the reduced density matrix (bipartite
entangled state $\rho_{ij}$ ) does not violate Bell-CHSH inequality. \\
Next we investigate whether this entangled state which doesn't
violate Bell's Inequality can still be used as a teleportation channel
with a fidelity greater than $\frac{2}{3}$ . However we find that
result obtained from this investigation is negative.\\
We recall the eigenvalues of the matrix
$U=C^t(\rho_{ij})C(\rho_{ij})$ to be as $u_1=u_2=\frac{4}{N^2}$,
$u_3=\frac{(N-4)^2}{N^2}$. Thus the teleportation fidelity
$F_{max}$ [8] becomes,
\begin{eqnarray}
F_{max}=\frac{1}{2}(1+\frac{1}{3}[\sqrt{u_1}+\sqrt{u_2}+\sqrt{u_3}])=
\frac{2}{3}.
\end{eqnarray}
Hence the state cannot be used as a teleportation channel since it does not overtake the classical fidelity.
In other words, in this section we conclude that any two qubit subsystem of a $N$ qubit W state can not be used as a teleportation
channel, in spite of being an entangled state without
violating Bell's inequality.  \\\\
\section{A random mixture
of a bipartite subsystem of a N-qubit W state and GHZ state:
Bell's Inequality and Teleportation} In this section we 
construct a state which is a classical mixture of the
bipartite reduced density matrices obtained from a $N$-qubit W state
and a $N$ qubit GHZ state. Here we also present a detailed analysis of various
interesting features like detection of its entanglement from
Peres-Horodecki criterion [14,15], violation of Bell's inequality
[5,6], and teleportation fidelity [1,2,3,4] for this state. Our aim is to study this state as a quantum channel for
teleportation.\\
This two qubit state represented by the density matrix $\rho_{GHZ,W}$ can be explicitly written as
\begin{eqnarray}
&&\rho_{GHZ,W}=pTr_{N-2}(|W\rangle\langle W|)+(1-p)Tr_{N-2}(|GHZ
\rangle\langle
GHZ|){}\nonumber\\&&=p\frac{1}{N}[|01\rangle\langle 01
|+|10\rangle\langle 10 |+|01\rangle\langle 10|+|10\rangle\langle
01|+(N-2)|00\rangle\langle
00|]{}\nonumber\\&&+\frac{1-p}{2}[|00\rangle\langle
00|+|11\rangle\langle 11|]~~~~~~(0<p<1).
\end{eqnarray}
This state can be viewed as a kind of a convex combination of the separable density matrix $Tr_{N-2}(|GHZ\rangle\langle GHZ|)$ and an inseparable density matrix $Tr_{N-2}(|W\rangle\langle W|)$.
Here we investigate the inseparability of the two qubit density
operator $\rho_{GHZ,W}$ for different values of $N$. After
calculating the determinants $W_3$ and $W_4$, we obtain the
values as
\begin{eqnarray}
&&W_3=[p\frac{(N-2)}{N}+\frac{1-p}{2}]\frac{p^2}{N^2}{}\nonumber\\&&
W_4=\frac{p^2}{4N^3}\{2p(1-p)N(N-2)+(1-p)^2N^2-4p^2\}.
\end{eqnarray}
Since $0<p<1$ and $N>2$, we conclude from the above expression that $W_3>0$. But
we can not say the same for $W_4$. Now $W_4<0$, when
$2p(1-p)N(N-2)+(1-p)^2N^2-4p^2<0$.\\
Next we obtain a relationship between $N$ and $p$ under which
the classical mixture $\rho_{GHZ,W}$  satisfies Bell-CHSH
inequality ($M(\rho_{GHZ,W})<1$).\\
The eigenvalues of the matrix $U=C^t(\rho_{GHZ,W})C(\rho_{GHZ,W})$
are given by $u_1=u_2=\frac{4p^2}{N^2}$,
$u_3=\frac{(N-4p)^2}{N^2}$. Now from here two possibilities can
arise.\\
\textbf{Case 1:} $u_1=u_2\geq u_3$.\\
This will happen when $p \in [\frac{N}{6},\frac{N}{2}]$. So in
this case, the largest two out of the three eigenvalues are
$u_1$,$u_2$, so $M(\rho_{GHZ,W})=u_1+u_2=\frac{8p^2}{N^2}<1$ for $(N>2)$.\\
\textbf{Case 2:} $u_3>u_1=u_2$.\\ This will happen when
$p\leq\frac{N}{6},p\geq\frac{N}{2}$. So in this case the largest two out of the three
eigenvalues are $u_1$,$u_3$,
$M(\rho_{GHZ,W})=u_1+u_3=\frac{4p^2}{N^2}+\frac{(N-4p)^2}{N^2}$.\\
Quite similar to the previous cases, next we find out a kind of relationship between $N$ and $p$ for which $\rho_{GHZ,W}$ can be used as an efficient teleportation channel.
Considering the same set of eigenvalues
$u_1=u_2=\frac{4p^2}{N^2}$, $u_3=\frac{(N-4p)^2}{N^2}$, the
teleportation fidelity is given
by\\
\textbf{Case1:} When $p\leq\frac{N}{4}$.
\begin{eqnarray}
F_{max}=\frac{2}{3}.
\end{eqnarray}
\textbf{Case2:} When $p>\frac{N}{4}$.
\begin{eqnarray}
F_{max}=\frac{1}{2}+\frac{8p-N}{6N}.
\end{eqnarray}
Now, interestingly $F_{max}>\frac{2}{3}$ when $p>\frac{N}{4}$ i.e
$p \in (\frac{N}{4},1)$ . This is because the order of the terms in the enumerator changes in either of the two cases.
This is an interesting result.  This
clearly indicates that when $N=3$, the state considered by us can be used as an efficient
teleportation channel. \\
So in this section we have analyzed three different
possibilities. These are (1) detection of its entanglement, (2)
non  violation of Bell's inequality, and (3) teleportation fidelity
greater than $\frac{2}{3}$. Interestingly we find out various
conditions involving $N$ and $p$ for which all of these above three
possibilities can be regarded as realities.
\section{Bell's Inequality and Teleportation for N=3,4,5}
Here we explicitly give a detailed analysis for $N=3,4,5$, and
obtain restrictions on the classical probability of mixing for each of these three cases. We find out the range of $p$ within which the state $\rho_{GHZ,W}$ can be treated as an
entangled state that does not violate Bell-CHSH inequality. Not only that we also find out that for $p\in (.75,1]$ and $N=3$, this state can be used as a teleportation channel. \\
In the previous section, we have employed Peres-Horodecki
criterion to detect the amount of entanglement in the state
$\rho_{GHZ,W}$. To show the density matrix $\rho_{GHZ,W}$ to be
entangled we have to show at least one of these determinants $W_3$ and
$W_4$ as a negative quantity. We have already found $W_3$ to be positive
for all values of $N$ and $p$. However we have obtained an inequality for which
$W_4$ is going to be negative. This inequality is given by,
\begin{eqnarray}
2p(1-p)N(N-2)+(1-p)^2N^2-4p^2<0.
\end{eqnarray}
In the following we investigate this inequality for $N=3,4,5$
respectively.\\
\textbf{Case 1 :} $N=3$\\
In this case when $N=3$, the reduced form of the inequality (9) is given by
\begin{eqnarray}
p^2+12p-9>0.
\end{eqnarray}
Here we see that for $p>.708$ , $W_4$ is going to be negative.\\
\textbf{Case 2 :} $N=4$\\
In this case when $N=4$, the reduced form of the inequality (9) is given by
\begin{eqnarray}
p^2+4p-4>0.
\end{eqnarray}
In this case for $p>.828$,  $W_4$ is going to be negative.\\
\textbf{Case 3 :} $N=5$\\
Now here when $N=5$, the inequality involving p is given by
\begin{eqnarray}
9p^2+20p-25>0.
\end{eqnarray}
This will hold when $p>.891$\\
From each of these thre cases we find that the state $\rho_{W,GHZ}$ is entangled for very
small range of $p$. \\
Thereafter we investigate whether this density matrix which is
entangled for very small range of $p$ is satisfying Bell's
inequality or not.\\
\textbf{Case 1:} When $N=3$, the density matrix $\rho_{W,GHZ}$ is
entangled for $p>.708$ i.e $p\in (.708,1)$. Now in this range the
value of $M(\rho_{W,GHZ})$ is given by
$M(\rho_{W,GHZ})=\frac{8p^2}{9}<1$. This is because 
$u_1=u_2>u_3$ when $p \in [\frac{N}{6}, \frac{N}{2}]$. This also implies that for $N=3$, 
 $p$ lies in the range $[.5,1.5]$ i.e $p\in [.5,1]$. \\
\textbf{Case 2:} When $N=4$, the density matrix $\rho_{W,GHZ}$ is
entangled for $p>.828$ i.e $p\in (.828,1)$. Now in this range the value
of $M(\rho_{W,GHZ})$ is given by $M(\rho_{W,GHZ})=\frac{8p^2}{16}<1$. This is because
$u_1=u_2>u_3$ when $p \in [\frac{N}{6}, \frac{N}{2}]$. This also implies that for $N=4$, $p$ lies in the interval $[.66,1]$. \\
\textbf{Case 3:} When $N=5$, the density matrix $\rho_{W,GHZ}$ is
entangled for $p>.891$ i.e $p\in (.891,1)$. Since for $N=5$, 
$u_1=u_2>u_3$ holds when  $p\in [.83,2.5]$ i.e $p\in [.83,1]$, so we consider the eigenvalues $u_1$ and $u_2$. Hence the value
of $M(\rho_{W,GHZ})$ is given by $M(\rho_{W,GHZ})=\frac{8p^2}{25}<1$.\\
In the following table we give the range of $p$ for
which the state $\rho_{W,GHZ}$ is an entangled state. We also report that for $N=3$ and $.75<p\leq 1$, this state can act as a teleportation channel without violating Bell's inequality.\\\\\\
{\bf TABLE 1:}\\
\begin{tabular}{|c|c|c|c|c|}
\hline  N & $W_4<0$ (range of p) & $M(\rho_{W,GHZ})$ & $M(\rho_{W,GHZ})\leq 1$ & $F_{max}>\frac{2}{3}$\\
\hline 3 & $[p \in(.708,1)]$ & $\frac{8p^2}{9}$ & Yes &Yes $[p \in(.75,1)]$, No $[p \in(.708,.75)]$\\
\hline 4 & $[p \in(.828,1)]$ & $\frac{p^2}{2}$ & Yes & No \\
\hline 5 & $[p \in(.891,1)]$ & $\frac{8p^2}{25}$ & Yes & No \\
\hline
\end{tabular}\\\\
\section{Conclusion} To summarize, in this paper our basic motivation is to look out for entangled states which can act as a
teleportation channel without violating Bell's inequality. Here we have considered two
types of two qubit entangled states. The first one is a two qubit subsystem of a $N$-qubit
W state. However we have found that such a state can not act as a teleportation channel and they are not violating Bell's inequality. Next we have considerd a state which we have mentioned is a classical
mixture of a two qubit subsystem of a $N$ qubit W state and a $N$ qubit GHZ state. We also investigated whether the violation of Bell's inequality has any bearing on the ability of these states
to teleport efficiently. More specifically, the equations (7), (8) give us the teleportation fidelity  also help us in finding out the condition within which this
entangled state can act as a teleportation channel without violating Bell's inequality. We conclude by noting that for mixed entangled
states like $\rho_{W,GHZ}$, the violation of Bell's inequality is not a good indicator
of their ability to perform as quantum information processing tasks such as teleportation.

\section{Acknowledgement}
I. C acknowledges Prof C. G. Chakraborti for being the source of
inspiration in research work. I. C acknowledges Dr. S. Adhikari and
N. Ganguly for having useful discussions. I.C also acknowledges Dr. Anirban Basu for helping in some technical problems.
\section{Reference}
$[1]$ C. H. Bennett, G. Brassard, C. Crepeau, R. Jozsa, A. Peres, W. K. Wootters, Phys.Rev.Lett. 70, 1895 (1993).\\
$[2]$ N. Gisin, Phys.Lett. A 210, 157 (1996).\\
$[3]$ R. Horodecki, P. Horodecki and M. Horodecki, Phys. Lett. A
200, 340 (1995).\\
$[4]$ S. Massar and S. Popescu, Phys.Rev.Lett. 74, 1259 (1995).\\
$[5]$ S. Popescu, Phys.Rev.Lett. 72, 797 (1994).\\
$[6]$ J. F. Clauser, M. A. Horne, A. Shimony, R. A. Holt, Phys.Rev.Lett.
23, 80 (1969).\\
$[7]$ R. F. Werner, Phys.Rev. A 40, 4277 (1989).\\
$[8]$ R. Horodecki, M. Horodecki, P. Horodecki, Phys.Lett. A 222, 21
(1996).\\
$[9]$ S. Adhikari, N. Ganguly, I. Chakrabarty,
B. S. Choudhury, J. Phys. A: Math. Theor. 41, 415302 (2008).\\
$[10]$ V. N. Gorbachev, A. A. Rodichkina and A. I. Trubilko, Phys.
Lett. A 310, 339 (2003).\\
$[11]$ J. Joo, Y. J. Park, S. Oh and J. Kim, New Journal of
Physics. 5, 136 (2003).\\
$[12]$ P. Agrawal, A. K. Pati, Phys Rev A 74, 062320 (2006).\\
$[13]$ S. Adhikari, S. Roy, B. Ghosh, A. S. Majumdar, N. Nayak,
Teleportation via maximally and non-maximally entangled mixed
states, arXiv: 0812.3772.\\
$[14]$  A. Peres, Phys.Rev.Lett. 77 1413 (1996).\\
$[15]$ M. Horodecki, P. Horodecki, R. Horodecki, Phys.Lett. A 223. 1 (1996).\\

\end{document}